\documentclass[aps,pre,twocolumn,superscriptaddress,reprint]{revtex4-1}
\usepackage{graphicx, color}
\usepackage{amsmath, amssymb, amsfonts, mathrsfs}

\begin{document}

\title{Free-energy functionals of the electrostatic potential for Poisson-Boltzmann theory}
\author{Vikram Jadhao}
\affiliation{Department of Materials Science and Engineering, Northwestern University, Evanston, Illinois 60208, USA}
\author{Francisco J. Solis}
\affiliation{School of Mathematical and Natural Sciences, Arizona State University, Glendale, Arizona 85306, USA}
\author{Monica Olvera de la Cruz}
\email{m-olvera@northwestern.edu}
\affiliation{Department of Materials Science and Engineering, Northwestern University, Evanston, Illinois 60208, USA}

\begin{abstract}
In simulating charged systems, it is often useful to treat some ionic components of the system
at the mean-field level and solve the Poisson-Boltzmann (PB) equation 
to get their respective density profiles. 
The numerically intensive task of solving the PB equation at each step of the simulation 
can be bypassed using variational methods that treat the electrostatic potential 
as a dynamic variable. But such approaches require the access to a true free-energy functional; 
a functional that not only provides the correct solution of the PB equation upon extremization,
it also evaluates to the true free energy of the system at its minimum. 
Moreover, the numerical efficiency of such procedures is further enhanced if 
the free-energy functional is local and is expressed in terms of the electrostatic potential.
Existing PB functionals of the electrostatic potential, while possessing the local structure, are not 
free-energy functionals. We present a variational formulation with a local 
free-energy functional of the potential. In addition, we also construct a nonlocal
free-energy functional of the electrostatic potential. 
These functionals are suited for employment in simulation schemes based on the ideas of dynamical
optimization.
\end{abstract}

\maketitle

\section{Introduction}
Electrostatic interactions are important in determining the structure and physical properties of 
charged biological systems such as proteins, DNA and cell membranes \cite{perutz,honig,clapham}.
These interactions also play a crucial role in realizing and stabilizing the  self-assembly 
of soft materials, such as colloidal dispersions in solution \cite{levin1}, and 
the formation of many synthetic materials such as 
patterned nano-structures \cite{paco} and faceted shells \cite{vernizzi}.
Along with the charged macromolecules, a typical biological
or synthetic system is inhabited by other charged entities such as counterions, salt ions and bound charges 
induced in the molecules of the solvent, leading to an enormous number of degrees of freedom (DOF) in the 
associated theoretical model.
Direct simulation of such a model is challenging even for current computers, 
and therefore approximations are introduced in the model with the motive of 
capturing the effects generated by certain elements of the system without 
explicitly including those elements in the model system, thus reducing the DOF required to simulate.
One such approximation is the implicit solvent model, where the molecular 
structure of the solvent is ignored and the solvent is treated as a dielectric continuum. 
Subsequently, one introduces the concept of dielectric permittivity in the model to replace the effects of 
discrete bound charges associated with the molecules of the solvent, resulting in a tremendous reduction 
in the number of DOF. 

However, in many cases, the number of ions present in the solution can be very large, 
proliferating the number of DOF constituting the model system, and thus limiting 
the size of the systems that can be simulated. 
Thus, in addition to the approximation of an implicit solvent, one proceeds to reduce the DOF in the 
model by ignoring the discrete nature of some ionic species of the charged system and replacing
their effects by a smeared-out, smoothly varying density distribution. 
Generally, such an approximation works very well when applied to the relatively mobile and weakly-charged 
components of the system, such as the monovalent ions among a mixture of monovalent and multivalent
salt near an oil-water emulsion, or counterions in the case of highly charge-asymmetric 
colloidal self-assembly \cite{lowen}. Accordingly, one partitions the overall charged system into: 
strongly-charged species that are modeled as finite-sized, \emph{fixed} discrete charges 
represented by a charge density $\rho_{f}$; and 
weakly-charged \emph{mobile} components modeled by appropriately chosen continuum distributions. 

Suppose the system has $N$ charged components that are treated via a continuum approach, and let
$c_{j}$ be the density of the $j^{\textrm{th}}$ component.
Often, a mean-field formalism suffices to describe the distributions of the $N$ components 
and one such widely used approach is the Poisson-Boltzmann (PB) theory \cite{andelman}.
Under this formalism, the density of the mobile components is
assumed to follow a Boltzmann distribution determined by the local electrostatic potential $\psi$:
\begin{equation}\label{eq:cj}
c_{j} = C_{j}\,e^{-\beta q_{j}\psi},
\end{equation}
where $\beta$ is the inverse thermal energy, 
$q_{j}$ and $C_{j}$ are respectively the charge and 
the bulk (reservoir) concentration of the $j^{\textrm{th}}$ component; the bulk being the region
where the mean electrostatic potential vanishes.
The total charge density in the system thus becomes $\rho = \rho_{f} + \sum_{j=1}^{N}q_{j}c_{j}$, where
$c_{j}$ is given by Eq.~\eqref{eq:cj}. Realizing that the electrostatic potential $\psi$, which
determines the component densities via Eq.~\eqref{eq:cj}, must itself obey the Poisson equation 
corresponding to the charge density $\rho$, we arrive at the well known Poisson-Boltzmann equation (PBE):
\begin{equation}\label{eq:pbe}
\nabla\cdot\left(\frac{\epsilon\nabla\psi}{4\pi}\right) 
+ \rho_{f} + \sum_{j=1}^{N}q_{j}C_{j}e^{-\beta q_{j}\psi} = 0,
\end{equation}
where $\epsilon(\mathbf{r})$ is the dielectric permittivity.

The PBE clearly reflects the model used to represent the complicated charged system:
$\rho_{f}$ contains the charged entities that are treated as discrete, finite-sized objects;
$\epsilon(\mathbf{r})$ encodes, in an approximate way, 
the information about the bound charges present on the solvent molecules; 
and the third term on the left hand side is the result of treating the ions belonging 
to some species as point particles, their smeared-out density being approximated by 
the PB theory. The solution of the PBE gives the equilibrium electrostatic potential. 
In general, for the simulation of systems represented via the above model, 
we are required to solve the PBE at each simulation step in order to obtain the forces 
required to propagate the configurational degrees of freedom. 
This leads to simulations that are very slow and costly as, at each step,
a very accurate solution of the PBE is needed to ensure proper energy conservation. 

Instead of solving the PBE, the electrostatic potential at equilibrium can also be obtained 
by appealing to the universal physical principle that 
the true equilibrium potential is the one that minimizes the system's free energy.
Accordingly, the PB theory is often recast as a variational problem where a suitable functional of 
the electrostatic potential is extremized to obtain the equilibrium potential.
In this alternative picture, the earlier assumption that mobile ions obey 
Boltzmann distribution translates into approximating the entropy of these ions to be that of an ideal gas.
With regards to the simulations of charged systems, the variational treatment opens the 
possibility of performing a simultaneous optimization of electrostatic and conformational 
degrees of freedom, offering a huge advantage over directly solving the above differential equation
at each step of the simulation. 
However, these benefits are only accessible if the variational principle is based on 
a true free-energy functional of the electrostatic potential, that is, 
a functional that evaluates to the true free energy of the system at its minimum. 

Unfortunately, a free-energy functional of the electrostatic potential whose minimizer satisfies 
the PBE does not exist in the literature. The standard PB functional of the potential \cite{honig1,radke},
as correctly noted by several authors \cite{fogolari,che,maggs1}, is not a free-energy functional; 
it \emph{maximizes} to the true free energy of the system. 
Existing PB free-energy functionals employ either expensive vector variables \cite{kung,maggs1}, 
such as the electric field $\mathbf{E}$ or the displacement field $\mathbf{D}$, which require a 
three-dimensional vectorial specification; or charge densities \cite{fogolari,che,prateek} 
which end up producing functionals involving nonlocal (long-range) interactions 
making the associated numerical minimization inefficient.
We note the existence of similar free-energy functionals for the case of linearized PBE \cite{van_roij,knott}. 
In view of this, we seek a local PB free-energy functional of the electrostatic potential, thus
combining the desirable features of locality, convexity, true equilibrium free energy,
and the use of a scalar field variable. 

In this article, we present a variational formulation that 
produces two PB free-energy functionals employing the electrostatic potential as their sole variational field.
One of these functionals is a local functional. 
While both our functionals are suited for employment 
in simulations carried out using dynamical optimization methods \cite{lowen,rottler-maggs,car-parrinello}, 
we envision the local functional in particular to be an excellent candidate to 
realize the possibility of simultaneous propagation of electrostatic and conformational degrees of 
freedom via the aforementioned optimization schemes. 

This paper is organized as follows. In Sec.~\ref{sec:fnals}, we present the free-energy functionals of 
the electrostatic potential whose minimizers satisfy the PBE. 
Section \ref{sec:varform} presents the variational formulation
that produces these functionals. In Sec.~\ref{sec:comparison}, we compare our local functional with the 
standard PB functional and use an example of a symmetric electrolyte to highlight the differences. 
Some concluding remarks are made in Sec.~\ref{sec:conclusion} and we end by providing proofs
related to the extremal properties of our functionals in 
Appendix \ref{app:proofs}.

\section{Poisson-Boltzmann Free-Energy Functionals of Potential}\label{sec:fnals}
The PB free-energy functionals of potential that we produce have the following basic form: 
each functional represents the free energy of the charged system constrained by 
the fundamental equation that the model system must obey, namely the PBE.
The key aspect that separates our functionals from other PB functionals of potential 
\cite{honig1,radke, fogolari} is the form of the constraint that is enforced by means of Lagrange multipliers.
Our constraint expressions endow the functionals with the desired features of true
equilibrium free energy and convexity. These constraint forms arise naturally out of our 
variational formulation which involves, as a key step, 
the recasting of the PBE into a recursive relation for the potential. 
A similar variational formulation was introduced by us recently \cite{jso1,jso2} wherein 
we recast the Poisson equation in a recursive form to construct a family of true energy functionals 
for electrostatics in heterogeneous media.

We begin by introducing some notations. Gaussian units will be used throughout. 
We define the function $h$ as 
\begin{equation}\label{eq:h}
h(\psi) = \sum_{j=1}^{N}q_{j}C_{j}e^{-\beta q_{j}\psi},
\end{equation}
and also define its inverse $h^{-1}$ via the following equation:
\begin{equation}\label{eq:hinv}
h^{-1}(h(\psi)) = \psi.
\end{equation}
The Green's function in free space is denoted by $G_{\mathbf{r},\mathbf{r'}}$ and we recall that it 
satisfies the relation
\begin{equation}\label{eq:G}
\nabla^{2}G_{\mathbf{r},\mathbf{r'}} = -4\pi\delta_{\mathbf{r},\mathbf{r'}},
\end{equation}
where $\delta_{\mathbf{r},\mathbf{r'}}$ is the Dirac-delta function. 

Our local PB free-energy functional reads as
\begin{equation}\label{eq:lpbef}
\begin{split}
\mathscr{K}_{\textrm{L}}[\psi]=&
\int\left(\frac{\epsilon\left|\nabla\psi\right|^{2}}{8\pi}
-\frac{1}{\beta}\sum_{j=1}^{N}C_{j}e^{-\beta q_{j}\psi}
\!\left(\beta q_{j}\psi + 1\right)\right)\!d\mathbf{r}\\
&+\int\Psi_{\textrm{L}}
\left(\nabla\cdot\frac{\epsilon}{4\pi}\nabla\psi 
+ \rho_{f} + h\left(\psi\right)\right) d\mathbf{r},
\end{split}
\end{equation}
where $\Psi_{\textrm{L}}$ is given by
\begin{equation}\label{eq:PsiL}
\begin{split}
\Psi_{\textrm{L}} = h^{-1}
\left(- \nabla\cdot\frac{\epsilon}{4\pi}\nabla\psi - \rho_{f}\right).
\end{split}
\end{equation}
Our nonlocal PB free-energy functional is:
\begin{equation}\label{eq:nlpbef}
\begin{split}
\mathscr{K}_{\textrm{NL}}[\psi]=&
\int\left(\frac{\epsilon\left|\nabla\psi\right|^{2}}{8\pi}
-\frac{1}{\beta}\sum_{j=1}^{N}C_{j}e^{-\beta q_{j}\psi}
\!\left(\beta q_{j}\psi + 1\right)\right)\!d\mathbf{r}\\
&+\int\Psi_{\textrm{NL}}
\left(\nabla\cdot\frac{\epsilon}{4\pi}\nabla\psi 
+ \rho_{f} + h\left(\psi\right)\right)d\mathbf{r},
\end{split}
\end{equation}
where $\Psi_{\textrm{NL}}$ is given by
\begin{equation}\label{eq:PsiNL}
\begin{split}
\Psi_{\textrm{NL}} = \int G_{\mathbf{r},\mathbf{r'}}
\left(\nabla\cdot\big(\chi_{\mathbf{r'}}\nabla\psi_{\mathbf{r'}}\big) + \rho_{f}(\mathbf{r'})+ h\big(\psi_{\mathbf{r'}}\big) \right)d\mathbf{r'},
\end{split}
\end{equation} 
and $\chi$ is the susceptibility related to the permittivity $\epsilon$ by the relation 
$\epsilon = 1 + 4\pi\chi$.

Extremizing either $\mathscr{K}_{\textrm{L}}$ or $\mathscr{K}_{\textrm{NL}}$ with respect to $\psi$ 
leads to Eq.~\eqref{eq:pbe}, the Poisson-Boltzmann equation. 
Also, both $\mathscr{K}_{\textrm{NL}}$ and $\mathscr{K}_{\textrm{L}}$ 
retain a simple interpretation at equilibrium, as owing to this extremum condition, 
the constraint term in either functional vanishes and the leftover term becomes the equilibrium free energy.
Furthermore, either functional becomes a minimum at its extremum 
(see Appendix \ref{app:proofs} for the proofs of the minimum property).
As is evident from Eqs.~\eqref{eq:lpbef} and \eqref{eq:nlpbef}, 
the only part where the functionals $\mathscr{K}_{\textrm{L}}$ and $\mathscr{K}_{\textrm{NL}}$ differ
is in the form of the Lagrange multiplier that enforces the constraint of PBE. 
While $\Psi_{\textrm{L}}$ endows the functional   
$\mathscr{K}_{\textrm{L}}$ with the feature of locality, 
the functional $\mathscr{K}_{\textrm{NL}}$ is nonlocal due to the form of the 
Lagrange multiplier $\Psi_{\textrm{NL}}$. It is important to note that the functional $\mathscr{K}_{\textrm{L}}$
combines all of the desired features of locality, convexity, true free energy and the use of a scalar 
field variable. 

The construction of $\mathscr{K}_{\textrm{L}}[\psi]$ hinges on the fact that the inverse
function $h^{-1}$ is known, which in some cases will only be available numerically. 
We point out that the functional produced by Maggs in Ref.~\onlinecite{maggs1} also requires a similar 
reciprocal transform. In many interesting cases, $h^{-1}$ is analytically available, examples include the
important system of symmetric two-component electrolyte for which the specific expression of 
the functional $\mathscr{K}_{\textrm{L}}$ will be provided in Sec.~\ref{sec:comparison}. 
We note that our variational formulation also 
produces local free-energy functionals that do not require any inverse or reciprocal transforms, 
but we discard these functionals in favor of $\mathscr{K}_{\textrm{L}}[\psi]$ due to the more symmetric
structure of the latter. We elaborate more on this last point in Sec.~\ref{sec:otherpb}. 

It is clear from Eqs.~\eqref{eq:lpbef} and \eqref{eq:PsiL} that for the functional 
$\mathscr{K}_{\textrm{L}}$ to be well defined, $\Psi_{\textrm{L}}$ and hence the inverse function 
$h^{-1}(- \nabla\cdot\frac{\epsilon}{4\pi}\nabla\psi - \rho_{f})$ must 
not assume infinite or imaginary values, and therefore care must be taken that the argument of 
$h^{-1}$ lies within the set of values for which the inverse function is well behaved. 
In the case of a potential $\psi$ that satisfies the PBE,
$- \nabla\cdot\frac{\epsilon}{4\pi}\nabla\psi - \rho_{f} = h(\psi)$,
implying that the Lagrange multiplier 
$\Psi_{\textrm{L}} = h^{-1}(- \nabla\cdot\frac{\epsilon}{4\pi}\nabla\psi - \rho_{f}) = \psi$
is always well behaved at equilibrium configurations.
During the course of the simulation, wherein $\mathscr{K}_{\textrm{L}}$ 
is optimized on-the-fly, the electrostatic potential that results via the optimization is expected 
to fluctuate around the instantaneous exact solution of the PBE. 
Thus, care must be taken that only those deviations from the exact solution are allowed
which keep the function $h^{-1}$ well behaved. 
In other words, the optimization (fictitious) dynamics which ``moves'' the function $\psi$ in conjunction 
with the update of the charge configuration must ensure that the potential has not 
sprung much ahead or lagged way behind the exact solution of the PBE. 
Accordingly, one chooses the simulation timestep and the constituents of the fictitious sub-system 
that forms the representation of the potential \cite{car-parrinello,madden,jso1}.

\section{Variational Formulation}\label{sec:varform}
In the first part of this section, we present the variational formulation that produces the functionals
$\mathscr{K}_{\textrm{L}}$ and $\mathscr{K}_{\textrm{NL}}$.
In the next subsection we show how PB free-energy functionals, 
different from $\mathscr{K}_{\textrm{L}}$ and $\mathscr{K}_{\textrm{NL}}$,
can be derived using our variational principle.

\subsection{Derivation of $\mathscr{K}_{\textrm{L}}$ and $\mathscr{K}_{\textrm{NL}}$}
\label{sec:derivation}
We begin by writing the free energy of the charged system in the form of the functional:
\begin{equation}\label{eq:free}
\begin{split}
\mathscr{K}[\mathbf{E},c_{j}] =&
\frac{1}{8\pi}\int \epsilon\left|\mathbf{E}\right|^{2} d\mathbf{r}\\
&+ \frac{1}{\beta}\int\sum_{j=1}^{N}\left(c_{j}\textrm{ln}\left(c_{j}\Lambda_{j}^{3}\right) - c_{j}\right) d\mathbf{r}\\
&- \int\sum_{j=1}^{N}\mu_{j}c_{j}\, d\mathbf{r},
\end{split}
\end{equation}
where $\Lambda_{j}$ and $\mu_{j}$ are respectively the deBroglie wavelength and chemical potential 
associated with the mobile ions of the $j^{\textrm{th}}$ component. 
In the above equation, the first term on the right hand side is the electrostatic energy
and the next two terms summarize the thermodynamic contribution to the free energy, 
where the approximation that the entropy of the mobile ions is equal to that of ideal gas particles
is employed to write the second term. 
We now introduce Gauss's law as a constraint to the above free-energy functional \cite{kung,baptista,maggs_cite}:
\begin{equation}\label{eq:in_const}
\begin{split}
\mathscr{K}[\mathbf{E},c_{j},\psi] =& 
\frac{1}{8\pi}\int \epsilon\left|\mathbf{E}\right|^{2} d\mathbf{r} \\
&+ \frac{1}{\beta}\int\sum_{j=1}^{N}\left(c_{j}\textrm{ln}\left(c_{j}\Lambda_{j}^{3}\right) - c_{j}\right) d\mathbf{r} \\ 
&- \int\sum_{j=1}^{N}\mu_{j}c_{j}\, d\mathbf{r} \\
&-\int \psi
\left(\nabla \cdot \left(\frac{\epsilon\mathbf{E}}{4\pi}\right) -
\rho_{f} - \sum_{j}q_{j}c_{j}\right) d\mathbf{r}.
\end{split}
\end{equation}
We note that the Lagrange multiplier $\psi$ used to enforce the constraint in the above equation
will turn out to be the mean-field electrostatic potential at equilibrium.
Also, we treat the fixed charge density $\rho_{f}$ as a parameter field and so we consider the above 
expression to be a functional of three variable fields: $\mathbf{E}$, $c_{j}$, and $\psi$. 
Moving forward, our goal is to eliminate all variables in favor of $\psi$, the desired 
variational field. 

Taking variations of $\mathscr{K}[\mathbf{E},c_{j},\psi]$ with respect to $\mathbf{E}$ and 
$c_{j}$ leads to the following set of equations,
\begin{equation}\label{eq:pbve}
\delta\mathbf{E} : \qquad \mathbf{E} = -\nabla\psi,
\end{equation}
\begin{equation}\label{eq:pbvcj}
\delta c_{j} : \qquad c_{j} = \frac{e^{\beta\mu_{j}}}{\Lambda_{j}^{3}}e^{-\beta q_{j}\psi}
= C_{j}e^{-\beta q_{j}\psi}.
\end{equation}
In Eq.~\eqref{eq:pbvcj}, we arrive at the second equality by absorbing 
the terms containing $\Lambda_{j}$ and 
$\mu_{j}$ into $C_{j}$, the latter becoming the bulk concentration, where the bulk is defined
as the region where the potential $\psi$ vanishes. 
From Eq.~\eqref{eq:pbve} we recover that the curl of the electric field must vanish 
(Maxwell's second equation). 
Equation \eqref{eq:pbvcj} implies that the concentration of the $j^{\textrm{th}}$ 
mobile component assumes a Boltzmann distribution corresponding to energy $q_{j}\psi$.
At this stage, employing Eqs.~\eqref{eq:pbve} and \eqref{eq:pbvcj}, 
we can eliminate $\mathbf{E}$ and $c_{j}$ from Eq.~\eqref{eq:in_const} in favor of $\psi$
and obtain a functional of the desired variational field. While this functional
does produce the correct mean-field potential upon extremization, 
it becomes a maximum, not a minimum, at its extremum. 
In fact, this functional is the standard PB functional 
found in the literature \cite{honig1,radke,fogolari,maggs1}. 
We elaborate more on the comparison between the standard PB functional and 
functionals produced by us in Sec.~\ref{sec:comparison}.

To arrive at the true free-energy functional of the electrostatic potential, 
as a first step, we resist substitution at this stage and instead 
take the un-utilized variation of $\mathscr{K}[\mathbf{E},c_{j},\psi]$ with respect to $\psi$, obtaining
\begin{equation}\label{eq:pbvpsi}
\delta\psi : \qquad 
\nabla \cdot \left(\frac{\epsilon\mathbf{E}}{4\pi}\right) - 
\rho_{f} - \sum_{j} q_{j}c_{j} = 0.
\end{equation}
In the above equation we can substitute $\mathbf{E}$ and $c_{j}$ in terms of $\psi$ using 
Eqs.~\eqref{eq:pbve} and \eqref{eq:pbvcj} respectively, to obtain
\begin{equation}\label{eq:relationpsi}
\nabla \cdot \left(\frac{\epsilon\nabla\psi}{4\pi}\right) + \rho_{f} + h(\psi) = 0,
\end{equation}
where we have deployed the notation $h$ defined in Eq.~\eqref{eq:h}.
Equation \eqref{eq:relationpsi} is identical to Eq.~\eqref{eq:pbe}, the PB equation. 
We are now at the key step of the derivation where in addition to eliminating $\mathbf{E}$ and $c_{j}$ from 
Eq.~\eqref{eq:in_const}, we employ the above equation to construct a different functional of $\psi$. The first step in this
process involves recasting Eq.~\eqref{eq:relationpsi} as a recursive relation for $\psi$ (see Eq.~\eqref{eq:lpsiinverse} below).
There are obviously many different ways to recast the above equation into a recursive relation for 
$\psi$ and depending on the particular recursive relation employed, we obtain the local or the nonlocal 
functional.

We begin with the manipulation of Eq.~\eqref{eq:relationpsi} that leads to the local functional 
$\mathscr{K}_{\textrm{L}}$.
This particular recasting begins by first writing Eq.~\eqref{eq:relationpsi} as 
\begin{equation}\label{eq:step1}
h(\psi) = - \nabla \cdot \left(\frac{\epsilon\nabla\psi}{4\pi}\right) - \rho_{f}.
\end{equation}
Using the definition of the inverse function $h^{-1}$ given in Eq.~\eqref{eq:hinv},
the above equation can be transformed into 
\begin{equation}\label{eq:lpsiinverse}
\psi = h^{-1}\left(- \nabla \cdot \left(\frac{\epsilon\nabla\psi}{4\pi}\right) - \rho_{f}\right).
\end{equation}
Equation \eqref{eq:lpsiinverse} is a recursive relation involving $\psi$ and following Eq.~\eqref{eq:PsiL}
we identify the right hand side of this equation to be the Lagrange multiplier $\Psi_{\textrm{L}}$.
Substituting $\mathbf{E}$, $c_{j}$, and $\psi$ from Eqs.~\eqref{eq:pbve}, \eqref{eq:pbvcj}, and 
\eqref{eq:lpsiinverse} respectively, back into Eq.~\eqref{eq:in_const} leads to  the local functional
$\mathscr{K}_{\textrm{L}}$.
We note that instead of employing the inverse function $h^{-1}$, 
Maggs \cite{maggs1} uses another reciprocal form, namely the Legendre transform of a function $g(\psi)$, 
related to the function $h(\psi)$ via $dg/d\psi = -h$, to construct his functional 
with the vector displacement $\mathbf{D}$ as the basic variable.

As alluded to earlier, Eq.~\eqref{eq:relationpsi} can be recast in another form suitable for the 
construction of the nonlocal PB free-energy functional $\mathscr{K}_{\textrm{NL}}$. 
This particular recasting begins by using the relation $\epsilon=1+4\pi\chi$ to split the first term 
on the left hand side of Eq.~\eqref{eq:relationpsi} and rearranging terms, which leads to 
\begin{equation}
\frac{\nabla^{2}\psi}{4\pi} =  - \nabla\cdot\big(\chi\nabla\psi\big) -\rho_{f} - h\big(\psi\big).
\end{equation}
By employing the basic property of Green's function, namely Eq.~\eqref{eq:G}, 
we can transform the above differential equation into an ``inverse'' integral form:
\begin{equation}\label{eq:nlpsi}
\psi = \int G_{\mathbf{r},\mathbf{r'}}
\Big\{\nabla\cdot\big(\chi_{\mathbf{r'}}\nabla\psi_{\mathbf{r'}}\big) + \rho_{f}(\mathbf{r'}) + h\big(\psi_{\mathbf{r'}}\big) \Big\}
d\mathbf{r'}.
\end{equation}
Equation \eqref{eq:nlpsi} is a recursive relation involving $\psi$ and we promptly identify the 
right hand side of this equation to be the Lagrange multiplier $\Psi_{\textrm{NL}}$.
Finally, the substitution of $\mathbf{E}$, $c_{j}$, and $\psi$ from Eqs.~\eqref{eq:pbve}, 
\eqref{eq:pbvcj}, and \eqref{eq:nlpsi} respectively, back in the functional of 
Eq.~\eqref{eq:in_const} leads to the nonlocal functional $\mathscr{K}_{\textrm{NL}}$.

\subsection{Other Poisson-Boltzmann Free-Energy Functionals}\label{sec:otherpb}
We note that $\mathscr{K}_{\textrm{L}}$ and $\mathscr{K}_{\textrm{NL}}$ are not the only two free-energy 
functionals that can be constructed via the above described variational formulation. 
By recasting Eq.~\eqref{eq:relationpsi} into alternative recursive relations, 
functionals that differ from those derived above can be constructed. For example, recalling that 
$h$ is a sum of charge densities for all mobile components present in the system, we consider 
splitting this function as $h(\psi) = h_{-}(\psi) + h_{+}(\psi)$, where $h_{-}$ 
includes the sum over only those component densities that describe negatively charged ions and 
$h_{+}$ is the sum over the component densities that represent only positively charged ions. 
Employing this splitting, we can transform Eq.~\eqref{eq:relationpsi} into a relation:
\begin{equation}\label{eq:lpsinew}
\psi = h_{+}^{-1}\left(- \nabla \cdot \left(\frac{\epsilon\nabla\psi}{4\pi}\right) - \rho_{f} 
- h_{-}(\psi)\right),
\end{equation}
where $h_{+}^{-1}$, is the inverse function defined by the relation $h_{+}^{-1}(h_{+}(\psi)) = \psi$.
The above equation provides a recursive relation involving $\psi$ that is
clearly different from the previous two relations expressed in 
Eqs.~\eqref{eq:lpsiinverse} and \eqref{eq:nlpsi}.
Substituting $\mathbf{E}$, $c_{j}$, and $\psi$ from Eqs.~\eqref{eq:pbve}, \eqref{eq:pbvcj}, and 
\eqref{eq:lpsinew} respectively, back into Eq.~\eqref{eq:in_const} leads to 
a local PB free-energy functional of $\psi$. 
Similarly, a different recursive relation is obtained by switching the $+$ and $-$ subscripts on $h$ in 
Eq.~\eqref{eq:lpsinew}, which leads to yet another local PB free-energy functional.
Proofs that the functionals obtained using these alternate substitutions are free-energy functionals
are similar to the ones that appear in the Appendix \ref{app:proofs} at the end of this paper.

We can construct more local PB free-energy functionals by using other ways to 
morph Eq.~\eqref{eq:relationpsi} into recursive relations for $\psi$.
As another example, we show a set of free-energy functionals that do not involve any inverse 
functions (such as $h^{-1}$ or $h_{+}^{-1}$). 
Such functionals can be constructed by noting that the splitting of the charge density $h$ 
discussed in the previous paragraph can be envisioned at the level of a single component. 
Accordingly, in Eq.~\eqref{eq:relationpsi}, we separate out 
the charge density term corresponding to the $k^{\textrm{th}}$ mobile component from the 
$h(\psi)$ term, and rearrange, thus obtaining
\begin{equation}\label{eq:l1}
q_{k}C_{k}e^{-\beta q_{k}\psi} = -\nabla \cdot \left(\frac{\epsilon\nabla\psi}{4\pi}\right) - 
\rho_{f} - \left(h -  q_{k}C_{k}e^{-\beta q_{k}\psi}\right).
\end{equation}
Dividing both sides of Eq.~\eqref{eq:l1} by $q_{k}C_{k}$ and taking the logarithm of 
each side leads to the relation:
\begin{equation}\label{eq:lpsi}
\psi = \frac{-1}{\beta q_{k}} 
\textrm{ln}\left(\frac{-\nabla\cdot\frac{\epsilon}{4\pi}\nabla\psi - \rho_{f} - h(\psi) 
+ q_{k}C_{k}e^{-\beta q_{k}\psi}}{q_{k}C_{k}}\right).
\end{equation}
Eq.~\eqref{eq:lpsi} is yet another cast of Eq.~\eqref{eq:relationpsi} which, unlike the previous
recursive relations, does not involve the use of inverse function. 
Once again, substituting $\mathbf{E}$, $c_{j}$, and $\psi$ from 
Eqs.~\eqref{eq:pbve}, \eqref{eq:pbvcj}, and \eqref{eq:lpsi}
respectively, in Eq.~\eqref{eq:in_const} leads to a PB free-energy functional of $\psi$.

We also note that nonlocal PB free-energy functionals different from $\mathscr{K}_{\textrm{NL}}$ 
can be constructed by using recursive relations different from the one in Eq.~\eqref{eq:nlpsi}. 
Beginning with the recursive relation in Eq.~\eqref{eq:nlpsi}
and using a procedure analogous to the one introduced by us in Ref.~\onlinecite{jso2}, 
a whole family of nonlocal free-energy functionals can be obtained.
Finally, it is natural to ponder if there exist recursive relations that do not 
lead to a free-energy functional. 
In order to answer this specific question and other related ones, 
a deeper understanding of the underlying variational principle is needed, which includes 
a general analysis of the process of constructing functionals constrained by a 
fundamental equation via a Lagrange multiplier. 
We postpone such an analysis to a future publication.

\section{Comparison with the standard Poisson-Boltzmann functional}\label{sec:comparison}
We noted in Sec.~\ref{sec:derivation} that not all substitutions to eliminate field variables from
Eq.~\eqref{eq:in_const} in favor of $\psi$ lead to a free-energy functional. We observed that
$\mathbf{E}$ and $c_{j}$ can be eliminated from Eq.~\eqref{eq:in_const} using Eqs.~\eqref{eq:pbve}
and \eqref{eq:pbvcj}, thus leading to a functional with $\psi$ as the sole variable. This process
results in the functional
\begin{equation}\label{eq:stdpb}
\begin{split}
I[\psi]=&
\int\left(\frac{\epsilon\left|\nabla\psi\right|^{2}}{8\pi}
-\frac{1}{\beta}\sum_{j=1}^{N}C_{j}e^{-\beta q_{j}\psi}
\!\left(\beta q_{j}\psi + 1\right)\right)\!d\mathbf{r}\\
&+\int\psi
\left(\nabla\cdot\frac{\epsilon}{4\pi}\nabla\psi 
+ \rho_{f} + h\left(\psi\right)\right) d\mathbf{r},
\end{split}
\end{equation}
which upon extremization, 
singles out the correct potential but becomes a maximum at equilibrium \cite{fogolari}.
$I[\psi]$ is in fact the standard PB functional \cite{honig1,radke}, although in literature one generally finds this 
functional expressed in the following equivalent form \cite{fogolari,maggs1}:
\begin{equation}\label{eq:stdpb1}
\begin{split}
I[\psi]=&
\int\left(-\frac{\epsilon\left|\nabla\psi\right|^{2}}{8\pi}
+ \rho_{f}\psi -\frac{1}{\beta}\sum_{j=1}^{N}C_{j}e^{-\beta q_{j}\psi}\right)d\mathbf{r},
\end{split}
\end{equation}
which can be derived from the functional in Eq.~\eqref{eq:stdpb} by integration by parts and using the 
definition of $h$.

It is clear from Eq.~\eqref{eq:stdpb} that $I$, like $\mathscr{K}_{\textrm{L}}$, 
is a local functional. 
From Eqs.~\eqref{eq:lpbef} and \eqref{eq:stdpb}, 
we note that functionals $\mathscr{K}_{\textrm{L}}$ and $I$ share a common structure:
the expression for the free energy (the first term in either functional) is constrained by the 
PBE. The only, but crucial, difference between these functionals is the choice of the constraint
that is enforced by means of Lagrange multipliers. While $I$ employs the function $\psi$ 
for enforcing the PBE constraint, $\mathscr{K}_{\textrm{L}}$ uses $\Psi_{\textrm{L}}$, 
obtained as a result of interpreting PBE as a recursive relation, for the same purpose.
Employment of $\Psi_{\textrm{L}}$ removes the defect of non-convexity present in the 
functional $I$, while retaining other desirable features.

It is useful to continue this comparison by using the specific example of 
a symmetric two-component electrolyte. For this system,
$q_{1} = q$, $q_{2} = -q$ and 
$C_{1} = C_{2} = C$. It is easy to show from Eq.~\eqref{eq:h} that 
the function $h$ for this problem becomes 
\begin{equation}\label{eq:hqq}
h(\psi) = -2qC\,\textrm{sinh}(\beta q\psi).
\end{equation}
Also, it follows from the above equation that the function $h^{-1}$ can be obtained analytically, and we find it to be: 
\begin{equation}\label{eq:hinvqq}
h^{-1}(y) = \frac{1}{\beta q}\textrm{sinh}^{-1}\left(y/\left(-2qC\right)\right).
\end{equation}
Using Eq.~\eqref{eq:stdpb1}, we find the standard PB functional for this system to be 
\begin{equation}\label{eq:stdpbqq}
\begin{split}
I[\psi]=&
\int\left(-\frac{\epsilon\left|\nabla\psi\right|^{2}}{8\pi}
+ \rho_{f}\psi
-\frac{2C}{\beta}\textrm{cosh}(\beta q\psi)\right)d\mathbf{r}.
\end{split}
\end{equation}
Judging by the (negative) signs that precede the $|\nabla\psi|^{2}$ and $\textrm{cosh}(\beta q \psi)$ terms
in the above functional (and noting that $\epsilon, \beta,$ and $C$ are all positive quantities), 
it is evident that this functional is unbounded from below. 
It takes little effort to show that the second variation of the above functional at its extremum is 
\begin{equation}\label{eq:sdiqq}
\begin{split}
\delta^{2}I=&
-\int\frac{\epsilon}{4\pi}\left|\nabla\delta\psi_{\mathbf{r'}}\right|^{2}d\mathbf{r'}\\
&-2\beta q^{2}C\int \textrm{cosh}(\beta q \psi)
\left(\delta\psi_{\mathbf{r'}}\right)^{2}d\mathbf{r'},
\end{split}
\end{equation}
which by inspection is a strictly negative number implying that $I$ becomes a maximum at equilibrium. 
One might promptly suggest multiplying an overall negative sign to $I[\psi]$ in Eq.~\eqref{eq:stdpbqq} to 
remove the defect of concavity.
However, doing so, as can be readily checked, leads to the  
wrong value of the equilibrium free energy for the system. 
In fact, considering the form of our local functional for this system as given below,
it will be very hard to guess how must $I[\psi]$ of Eq.~\eqref{eq:stdpbqq} be modified in order to
convert it into a free-energy functional. 
It is therefore not a surprise that previous attempts at constructing free-energy functionals
for PB theory have instead ended up changing the field variable from $\psi$ to either the vector variables such as 
$\mathbf{D}$ or $\mathbf{E}$ \cite{maggs1,kung}, or charge densities \cite{che, prateek, fogolari}.

We now obtain the expression for the functional $\mathscr{K}_{\textrm{L}}$ in the case of the symmetric two-component
electrolyte. Employing Eqs.~\eqref{eq:hqq} and \eqref{eq:hinvqq} in Eq.~\eqref{eq:lpbef}, 
we find $\mathscr{K}_{\textrm{L}}$ to be
\begin{equation}\label{eq:lpbefqq}
\begin{split}
\mathscr{K}_{\textrm{L}}[\psi]=&
\int\frac{\epsilon\left|\nabla\psi\right|^{2}}{8\pi}d\mathbf{r}\\
&-\int\left(\frac{2C}{\beta}\textrm{cosh}(\beta q\psi)
- 2qC\psi \, \textrm{sinh}(\beta q\psi)\right)\!d\mathbf{r}\\
&+\int\frac{1}{\beta q}\textrm{sinh}^{-1}\left(\frac{1}{2qC}\left(
\nabla\cdot\frac{\epsilon\nabla\psi}{4\pi} + \rho_{f}\right)\right)\\
&\qquad\times\left(\nabla\cdot\frac{\epsilon}{4\pi}\nabla\psi 
+ \rho_{f} - 2qC\,\textrm{sinh}(\beta q \psi)\right) d\mathbf{r}.
\end{split}
\end{equation}
Unlike the standard PB functional for this system, it is not immediately obvious if 
the above functional is convex or concave. So we turn to Eq.~\eqref{eq:lsd4} in Appendix \ref{app:proofs},  
which provides the general expression for the second variation of $\mathscr{K}_{\textrm{L}}$ at its 
extremum. Evaluating this variation for the parameters associated with the symmetric 
two-component electrolyte, we obtain
\begin{equation}\label{eq:sdkqq}
\begin{split}
\delta^{2}\mathscr{K}_{\textrm{L}}[\psi] =&\,
3\int\frac{\epsilon}{4\pi}\left|\nabla\delta\psi_{\mathbf{r'}}\right|^{2}d\mathbf{r'}\\
&+ 2\beta q^{2}C\int \textrm{cosh}(\beta q \psi)
\left(\delta\psi_{\mathbf{r'}}\right)^{2}d\mathbf{r'}\\
&+ \int \frac{\left(\nabla\cdot(\epsilon/4\pi)\nabla\delta\psi_{\mathbf{r'}}\right)^{2}}
{\beta q^{2}C\textrm{cosh}(\beta q \psi)}d\mathbf{r'}.
\end{split}
\end{equation}
It is clear that the right hand side of the above equation is a strictly positive number, 
thus implying that $\mathscr{K}_{\textrm{L}}$ becomes a minimum at its extremum. 
As the comparison of Eqs.~\eqref{eq:sdiqq} and \eqref{eq:sdkqq} reveals, 
the use of a recursive form $\Psi_{\textrm{L}}$ (in place of $\psi$) in the constraint term in 
Eq.~\eqref{eq:lpbefqq} flips the signs associated with the integrals involving 
$|\nabla\delta\psi_{\mathbf{r'}}|^{2}$ and $\textrm{cosh}(\beta q \psi)$ terms from negative to 
positive and, in addition, leads to a term which is strictly non-negative. 

Judging by the complicated form of the functional $\mathscr{K}_{\textrm{L}}$ in Eq.~\eqref{eq:lpbefqq}, 
we believe that this functional can not be obtained via a trivial manipulation of the 
functional $I[\psi]$ of Eq.~\eqref{eq:stdpbqq}.
We do not, however, claim that one can not arrive at free-energy functionals with simpler forms as compared to
the functional appearing in Eq.~\eqref{eq:lpbefqq}. It would indeed be useful to find such functionals 
via the variational principle presented here or otherwise.
Finally, we note that in spite of some similarities between the functional in Eq.~\eqref{eq:lpbefqq} and the 
functional of variable $\mathbf{D}$ derived by Maggs in Ref.~\onlinecite{maggs1} 
(for example, both functionals contain the term $\textrm{sinh}^{-1}(\xi)$, where 
$\xi = (\nabla\cdot\frac{\epsilon\nabla\psi}{4\pi} + \rho_{f})/2qC = (\rho_{f} - \nabla\cdot\frac{\mathbf{D}}{4\pi})/2qC$),
these two functionals are different functionals and it is not possible to obtain one from another by a simple 
transformation like $\mathbf{D} \to -\epsilon\nabla\psi$.

\section{Conclusion}\label{sec:conclusion}
In this article, we presented a variational formulation that produces true free-energy functionals
of the electrostatic potential whose minimizers satisfy the Poisson-Boltzmann equation. 
With the construction of a local PB free-energy functional of potential we have shown that 
the advantages of convexity, true equilibrium free energy and locality can all be embedded in a functional 
of a scalar field, without the need to move to the more expensive vector field variables. 
While both PB functionals we produce are strong candidates for simulation techniques aimed at 
solving PB equation on-the-fly in conjunction with the update of other conformational degrees of freedom,
we envision our local functional in particular to be an ideal choice for the powerful 
local optimization procedures \cite{rottler-maggs}.

We have also shown the versatile nature of our variational formulation which, in addition 
to producing the functionals $\mathscr{K}_{\textrm{L}}$ and $\mathscr{K}_{\textrm{NL}}$, 
is capable of constructing many other PB free-energy functionals.
Our formulation also reveals that functionals such as the standard PB functional of potential, which 
are not true free-energy functionals, can be understood as arising from deficient forms of the constraint 
of PBE applied to the free energy of the system. 
In this light, we believe our formulation and the associated free-energy functionals provide a 
fresh look at the PB theory.
The central feature of our formulation, which is to recast the fundamental equation (in the present case, 
the PBE) in a suitable recursive form for the construction of a true energy functional, appears to be
a very general idea, and it is our goal in future to investigate if this idea can be employed
for the construction of similar functionals for other theories such as 
classical electrostatics \cite{jackson, schwinger} or classical mechanics. 

\section{Acknowledgements}
The authors thank Z. Yao and G. Vernizzi for many useful conversations.
V.J. was funded by the Department of Defense Research and Engineering (DDR\&E) 
and the Air Force Office of Scientific
Research (AFOSR) under Award No. FA9550-10-1-0167.
F.J.S. was funded by the National Science Foundation (NSF)
Grants No. DMR-0805330 and No. DMR-0907781.

\appendix
\section{Extremal behavior of $\mathscr{K}_{\textrm{L}}$ and $\mathscr{K}_{\textrm{NL}}$ }\label{app:proofs}
In this appendix, we examine the functionals $\mathscr{K}_{\textrm{L}}$ and $\mathscr{K}_{\textrm{NL}}$
at the point of their extremum. Specifically, we prove that these functionals 
become a minimum at their extremum. In this regard, we will show that the second variation of 
either functional is strictly positive at the extremum point. 

We begin by introducing some notations. 
We define a function $B(\psi)$ as
\begin{equation}\label{eq:B}
B(\psi) = \nabla \cdot\frac{\epsilon}{4\pi}\nabla\psi + \rho_{f} + h(\psi),
\end{equation}
and use the functional $F[\psi]$ to denote 
\begin{equation}\label{eq:F}
\begin{split}
F[\psi]=&
\int\left(\frac{\epsilon\left|\nabla\psi\right|^{2}}{8\pi}
-\frac{1}{\beta}\sum_{j=1}^{N}C_{j}e^{-\beta q_{j}\psi}
\!\left(\beta q_{j}\psi + 1\right)\right)\!d\mathbf{r}\\
&+\int\psi B(\psi)d\mathbf{r}.
\end{split}
\end{equation}
Also, from now onwards, unless otherwise stated explicitly, $\mathscr{K}$ denotes either of our
PB functionals ($\mathscr{K}_{\textrm{L}}$ or $\mathscr{K}_{\textrm{NL}}$) and 
$\Psi$ denotes either multiplier ($\Psi_{\textrm{L}}$ or $\Psi_{\textrm{NL}}$).

To make the derivations smoother, when employing integration by parts 
we will quietly render the resulting surface integrals void by invoking the Dirichlet boundary 
condition. Also, dummy variables are used in abundance 
and they will appear (disappear) without notice. 
To make equations less congested, we often omit the 
position variable dependence of functions such as $\psi$, and trust the reader 
to figure the associated variable from the context. However, when the context is not conclusive, we
will explicitly show the variable dependence.
We frequently employ the identity 
\begin{equation}\label{eq:id1}
\frac{\delta\psi(\mathbf{r'})}{\delta\psi(\mathbf{r})} = \delta_{\mathbf{r},\mathbf{r'}}.
\end{equation}
And finally, using the definition of $B$, we recognize that the Poisson-Boltzmann equation is simply
written as 
\begin{equation}\label{eq:B0}
B(\psi) = 0,
\end{equation}
which, as can be readily verified, is identical to the relation: $\psi = \Psi$. 
This observation is used often in our proofs to which we turn next.

Using Eq.~\eqref{eq:B}, we can write our functional $\mathscr{K}$ as
\begin{equation}\label{eq:pbef}
\begin{split}
\mathscr{K}[\psi]=&
\int\left(\frac{\epsilon\left|\nabla\psi\right|^{2}}{8\pi}
-\frac{1}{\beta}\sum_{j=1}^{N}C_{j}e^{-\beta q_{j}\psi}
\!\left(\beta q_{j}\psi + 1\right)\right)\!d\mathbf{r}\\
&+\int\Psi B(\psi) d\mathbf{r}.
\end{split}
\end{equation}
We add and subtract $\psi$ from the multiplier $\Psi$ in the above equation and 
employ the definition of $F$ given in Eq.~\eqref{eq:F} to obtain
\begin{equation}\label{eq:pbef1}
\begin{split}
\mathscr{K}[\psi]= F[\psi] +\int (\Psi - \psi) B(\psi) d\mathbf{r}.
\end{split}
\end{equation}
The above expression serves as the starting point to compute the first derivative of 
$\mathscr{K}$, which is
\begin{equation}\label{eq:fd1}
\begin{split}
\frac{\delta\mathscr{K}[\psi]}{\delta\psi}=& 
\frac{\delta F[\psi]}{\delta\psi} +
\int (\Psi - \psi) \frac{\delta B(\psi)}{\delta\psi} d\mathbf{r'}\\
&+\int B(\psi)\frac{\delta(\Psi - \psi)}{\delta\psi} d\mathbf{r'}.
\end{split}
\end{equation}
Using Eq.~\eqref{eq:F} and employing integration by parts and making abundant use of 
Eq.~\eqref{eq:id1}, we compute the functional derivative of $F$ to be
\begin{equation}\label{eq:dF}
\begin{split}
\frac{\delta F[\psi]}{\delta \psi}= \int B(\psi_{\mathbf{r'}})\delta_{\mathbf{r},\mathbf{r'}}d\mathbf{r'}.
\end{split}
\end{equation}
Substituting $\frac{\delta F[\psi]}{\delta \psi}$ from the above equation into 
Eq.~\eqref{eq:fd1} and simplifying leads to our final expression for the first derivative of $\mathscr{K}$:
\begin{equation}\label{eq:fd2}
\begin{split}
\frac{\delta\mathscr{K}[\psi]}{\delta\psi}=& 
\int (\Psi - \psi) \frac{\delta B(\psi)}{\delta\psi} d\mathbf{r'}
+\int B(\psi)\frac{\delta\Psi}{\delta\psi} d\mathbf{r'}.
\end{split}
\end{equation}
It is clear from the above equation that when $B(\psi)=0$, 
which is identical to $\psi = \Psi$ as observed before,
$\frac{\delta\mathscr{K}[\psi]}{\delta \psi}$ vanishes.
But, as argued above, $B(\psi)=0$ is precisely the PBE. 
In other words, the potential that satisfies the PBE also extremizes the functional $\mathscr{K}$.

We now compute the second derivative of $\mathscr{K}$ and examine its value at the point of extremum.
Carrying out the derivative of $\frac{\delta\mathscr{K}}{\delta\psi}$ given in Eq.~\eqref{eq:fd2}, 
we obtain
\begin{equation}\label{eq:sd1}
\begin{split}
\frac{\delta^{2}\mathscr{K}[\psi]}{\delta \psi^{2}}=&
\int (\Psi - \psi) \frac{\delta^{2} B(\psi)}{\delta\psi^{2}} d\mathbf{r'} + 
\int \frac{\delta(\Psi - \psi)}{\delta\psi_{\mathbf{r}}} \frac{\delta B(\psi)}{\delta\psi_{\mathbf{r''}}} d\mathbf{r'}\\ 
&+\int \frac{\delta B(\psi)}{\delta\psi_{\mathbf{r}}}\frac{\delta\Psi}{\delta\psi_{\mathbf{r''}}} d\mathbf{r'}
+\int B(\psi)\frac{\delta^{2}\Psi}{\delta\psi^{2}} d\mathbf{r'},
\end{split}
\end{equation}
where we employ $\frac{\delta^{2}}{\delta\psi^{2}}$ to denote
$\frac{\delta^{2}}{\delta\psi(\mathbf{r})\delta\psi(\mathbf{r''})}$ for the sake of brevity.

Recalling that at the extremum point, $B(\psi) = 0$, and equivalently, $\psi - \Psi = 0$, we find 
that the first and last terms in Eq.~\eqref{eq:sd1} vanish upon using these equalities, and the 
above expression for the second derivative reduces at the point of extremum to
\begin{equation}\label{eq:sd2}
\frac{\delta^{2}\mathscr{K}[\psi]}{\delta \psi^{2}}\Big\vert_{\textrm{e}}=
\int \frac{\delta(\Psi - \psi)}{\delta\psi_{\mathbf{r}}} \frac{\delta B(\psi)}{\delta\psi_{\mathbf{r''}}} d\mathbf{r'}
+\int \frac{\delta B(\psi)}{\delta\psi_{\mathbf{r}}}\frac{\delta\Psi}{\delta\psi_{\mathbf{r''}}} d\mathbf{r'},
\end{equation}
which on further simplification and subsequent integration against the variations
$\delta\psi(\mathbf{r})$ and $\delta\psi(\mathbf{r''})$ leads
to the following expression for the second variation of the functional:
\begin{equation}\label{eq:sd3}
\delta^{2}\mathscr{K}[\psi]\Big\vert_{\textrm{e}}=
2\int \delta\Psi\delta B(\psi) d\mathbf{r'}
-\int \delta\psi_{\mathbf{r'}}\delta B(\psi)d\mathbf{r'},
\end{equation}
where we employ the shorthand notations 
$\delta\Psi = \int\frac{\delta\Psi}{\delta\psi(\mathbf{r})}\delta\psi(\mathbf{r}) d\mathbf{r}$
and $\delta B = \int \frac{\delta B}{\delta\psi(\mathbf{r''})}\delta\psi(\mathbf{r''}) d\mathbf{r''}$.
The subscript e used in Eqs.~\eqref{eq:sd2} and \eqref{eq:sd3} indicates that the 
second derivative and variation are being evaluated at the point of extremum. 

Equation \eqref{eq:sd3} provides the expression for the second variation of 
the functional $\mathscr{K}$ evaluated at its extremum. 
We will now examine the value of this variation for the local and the nonlocal functional separately.
We begin with the local functional. To compute 
$\delta^{2}\mathscr{K}_{\textrm{L}}[\psi]$ using 
Eq.~\eqref{eq:sd3}, we require the derivative of $\Psi_{\textrm{L}}(\psi(\mathbf{r'}))$ 
with respect to $\psi(\mathbf{r})$.
Remembering the definition of $\Psi_{\textrm{L}}$ from Eq.~\eqref{eq:PsiL}, we have
\begin{equation}\label{eq:lsd1}
\begin{split}
\frac{\delta\Psi_{\textrm{L}}(\mathbf{r'})}{\delta\psi_{\mathbf{r}}} \,=\, &
\frac{\delta h^{-1}\left(y\right)}{\delta\psi_{\mathbf{r}}}\\
\,=\, &\frac{\delta h^{-1}(y)}{\delta y}\cdot\frac{\delta y}{\delta\psi_{\mathbf{r}}}\\
\,=\, &\left(\frac{\delta h}{\delta \psi_{\mathbf{r'}}}\Big\vert_{\psi_{\mathbf{r'}}=h^{-1}(y)}\right)^{-1}
\cdot\frac{\delta y}{\delta\psi_{\mathbf{r}}},
\end{split}
\end{equation}
where the function $y$ stands for 
\begin{equation}\label{eq:y}
y = - \nabla\cdot\frac{\epsilon}{4\pi}\nabla\psi_{\mathbf{r'}} - \rho_{f}(\mathbf{r'}).
\end{equation}
Employing the definitions of $h$ and $y$ given by Eqs.~\eqref{eq:h} and \eqref{eq:y} respectively, 
Eq.~\eqref{eq:lsd1} becomes
\begin{equation}\label{eq:lsd2}
\begin{split}
\frac{\delta \Psi_{\textrm{L}}}{\delta\psi_{\mathbf{r}}}=&
\frac{1}{-\beta\sum_{j=1}^{N}C_{j}q_{j}^{2}e^{-\beta q_{j}h^{-1}(y)}}\left(-\nabla\cdot
\frac{\epsilon}{4\pi}\nabla\delta_{\mathbf{r},\mathbf{r'}}\right)\\
=&\frac{1}
{\beta\sum_{j=1}^{N}C_{j}q_{j}^{2}e^{-\beta q_{j}\Psi_{\textrm{L}}(\mathbf{r'})}}
\nabla\cdot\frac{\epsilon}{4\pi}\nabla\delta_{\mathbf{r},\mathbf{r'}},
\end{split}
\end{equation}
where we have used Eq.~\eqref{eq:id1} and remembered that $\Psi_{\textrm{L}} = h^{-1}(y)$ to obtain
the second equality above.
Since the above derivative needs to be evaluated at equilibrium, for which $\psi(\mathbf{r'}) = \Psi_{\textrm{L}}(\mathbf{r'})$,
we obtain
\begin{equation}\label{eq:lsd3}
\begin{split}
\frac{\delta \Psi_{\textrm{L}}}{\delta\psi_{\mathbf{r}}}\Big\vert_{\textrm{e}}
=\frac{1}
{\beta\sum_{j=1}^{N}C_{j}q_{j}^{2}e^{-\beta q_{j}\psi_{\mathbf{r'}}}}
\nabla\cdot\frac{\epsilon}{4\pi}\nabla\delta_{\mathbf{r},\mathbf{r'}}.
\end{split}
\end{equation}

From Eq.~\eqref{eq:B}, and using Eq.~\eqref{eq:id1}, we readily derive:
\begin{equation}\label{eq:dB}
\frac{\delta B(\psi_{\mathbf{r'}})}{\delta\psi_{\mathbf{r''}}} = 
\nabla\cdot\frac{\epsilon}{4\pi}\nabla\delta_{\mathbf{r''},\mathbf{r'}}
-\beta\sum_{j=1}^{N}C_{j}q_{j}^{2}e^{-\beta q_{j}\psi_{\mathbf{r'}}}\delta_{\mathbf{r''},\mathbf{r'}}.
\end{equation}
Using Eqs.~\eqref{eq:lsd3} and \eqref{eq:dB} in Eq.~\eqref{eq:sd3}, followed by integrating by parts 
and some simple algebra leads to the following equation:
\begin{equation}\label{eq:lsd4}
\begin{split}
\delta^{2}\mathscr{K}_{\textrm{L}}[\psi]\Big\vert_{\textrm{e}}=&\,
3\int\frac{\epsilon}{4\pi}\left|\nabla\delta\psi_{\mathbf{r'}}\right|^{2}d\mathbf{r'}\\
&+ \int\beta\sum_{j=1}^{N}C_{j}q_{j}^{2}e^{-\beta q_{j}\psi_{\mathbf{r'}}}
\left(\delta\psi_{\mathbf{r'}}\right)^{2}d\mathbf{r'}\\
&+ 2\int \frac{\left(\nabla\cdot(\epsilon/4\pi)\nabla\delta\psi_{\mathbf{r'}}\right)^{2}}
{\beta\sum_{j=1}^{N}C_{j}q_{j}^{2}e^{-\beta q_{j}\psi_{\mathbf{r'}}}}
d\mathbf{r'}.
\end{split}
\end{equation}

Since $\epsilon \ge 1$ and bulk concentrations are positive quantities, we find that each of 
the three integrals in Eq.~\eqref{eq:lsd4} is non-negative. Moreover, the first integral in 
the above equation is strictly greater than zero. Therefore, we have 
\begin{equation}
\delta^{2}\mathscr{K}_{\textrm{L}}[\psi]\Big\vert_{\textrm{e}} > 0,
\end{equation}
which completes the proof that the local functional $\mathscr{K}_{\textrm{L}}$ becomes a minimum 
when extremized. 

We now turn to the case of the nonlocal functional $\mathscr{K}_{\textrm{NL}}$.  
Using Eq.~\eqref{eq:sd3}, we now evaluate 
$\delta^{2}\mathscr{K}_{\textrm{NL}}[\psi]$ at the extremum point, 
the latter being again given by $B(\psi) = 0$, 
or equivalently $\psi - \Psi_{\textrm{NL}} = 0$. As is evident from
Eq.~\eqref{eq:sd3}, to proceed with this evaluation, we require the derivative of 
$\Psi_{\textrm{NL}}$ with respect to $\psi$. Note that in Eq.~\eqref{eq:sd3}, 
$\Psi_{\textrm{NL}}$ is a function of $\mathbf{r'}$ and its derivative is with respect to 
$\psi(\mathbf{r})$.
Remembering the definition of $\Psi_{\textrm{NL}}$ from Eq.~\eqref{eq:PsiNL}, we have
\begin{equation}\label{eq:nlsd1}
\begin{split}
\frac{\delta\Psi_{\textrm{NL}}}{\delta\psi_{\mathbf{r}}}
\,=\, \frac{\delta}{\delta\psi_{\mathbf{r}}} \int G_{\mathbf{r'},\mathbf{r''}} 
\left(B(\psi_{\mathbf{r''}}) -\frac{1}{4\pi}\nabla^{2}\psi_{\mathbf{r''}} \right) d\mathbf{r''},
\end{split}
\end{equation}
where, we have employed the definition of $B$ to simplify the expression for $\Psi_{\textrm{NL}}$.
Taking the derivative inside the integral and employing Eq.~\eqref{eq:id1} we obtain
\begin{equation}\label{eq:nlsd2}
\begin{split}
\frac{\delta\Psi_{\textrm{NL}}}{\delta\psi_{\mathbf{r}}} \,=\, &
\int G_{\mathbf{r'},\mathbf{r''}}
\left(\frac{\delta B(\psi_{\mathbf{r''}})}{\delta\psi_{\mathbf{r}}} -\frac{1}{4\pi}\nabla^{2}\delta_{\mathbf{r},\mathbf{r''}}\right)
d\mathbf{r''}.
\end{split}
\end{equation}
Using Eq.~\eqref{eq:nlsd2}, the integral in the first term on the right hand side 
of Eq.~\eqref{eq:sd3} becomes
\begin{equation}\label{eq:nlsd3}
\begin{split}
\int \delta\Psi_{\textrm{NL}}(\psi_{\mathbf{r'}}) \delta B(\psi_{\mathbf{r'}}) d\mathbf{r'} =&
\iint \delta B(\psi_{\mathbf{r'}}) G_{\mathbf{r'},\mathbf{r''}}\times\\
&\left(\delta B(\psi_{\mathbf{r''}}) -\frac{1}{4\pi}\nabla^{2}\delta\psi_{\mathbf{r''}}\right)
d\mathbf{r''}
d\mathbf{r'}.
\end{split}
\end{equation}

Before proceeding further, we introduce a shorthand $f(\mathbf{r'})$ which stands for 
\begin{equation}\label{eq:f}
f_{\mathbf{r'}} = \delta B(\psi_{\mathbf{r'}}) - \frac{1}{4\pi}\nabla^{2}\delta\psi_{\mathbf{r'}}.
\end{equation}
Adding and subtracting $\nabla^{2}\delta\psi_{\mathbf{r'}}/4\pi$ from 
$\delta B(\psi(\mathbf{r'}))$ in Eq.~\eqref{eq:nlsd3} and using 
the shorthand $f$ we can write Eq.~\eqref{eq:nlsd3} as
\begin{equation}\label{eq:nlsd4}
\begin{split}
\int \delta\Psi_{\textrm{NL}}(\psi_{\mathbf{r'}})\delta B(\psi_{\mathbf{r'}}) d\mathbf{r'} =&
\iint f_{\mathbf{r'}} G_{\mathbf{r'},\mathbf{r''}} f_{\mathbf{r''}}
\,d\mathbf{r''}d\mathbf{r'} +\\
&\iint\frac{1}{4\pi}\nabla^{2}\delta\psi_{\mathbf{r'}} 
G_{\mathbf{r'},\mathbf{r''}} f_{\mathbf{r''}}\,d\mathbf{r''}d\mathbf{r'}.
\end{split}
\end{equation}
We call the first and second double integrals in the above equation as $I_{1}$ and $I_{2}$
respectively. By inserting a $\delta$ function in 
the first double integral of Eq.~\eqref{eq:nlsd4},
and employing Eq.~\eqref{eq:G}, followed by a series of integration by parts,
$I_{1}$ can be written as 
\begin{equation}\label{eq:id2}
\begin{split}
I_{1}
=\int \left| \int \nabla_{\mathbf{r''}} G_{\mathbf{r'},\mathbf{r''}} f_{\mathbf{r''}} 
\,d\mathbf{r''} \right|^{2}d\mathbf{r'}.
\end{split}
\end{equation}
In the second double integral of Eq.~\eqref{eq:nlsd4}, by employing integration 
by parts, we can transfer the action of Laplacian
from $\delta\psi_{\mathbf{r'}}$ on to $G_{\mathbf{r'},\mathbf{r''}}$ and further simplify
to obtain
\begin{equation}\label{eq:nlsd5}
\begin{split}
I_{2}
&= \iint\delta\psi_{\mathbf{r'}} 
\frac{\nabla^{2}G_{\mathbf{r'},\mathbf{r''}}}{4\pi} f_{\mathbf{r''}}\,d\mathbf{r''}d\mathbf{r'}\\
&=-\iint\delta\psi_{\mathbf{r'}} 
\delta_{\mathbf{r'},\mathbf{r''}} f_{\mathbf{r''}}\,d\mathbf{r''}d\mathbf{r'}\\
&=-\int\delta\psi_{\mathbf{r'}}f_{\mathbf{r'}}\,d\mathbf{r'},
\end{split}
\end{equation}
where we have used Eq.~\eqref{eq:G} to get the second equality in the above equation and the final 
equality follows by carrying out the integral over $\mathbf{r''}$.
Substituting $I_{1}$ and $I_{2}$ from Eqs.~\eqref{eq:id2} and \eqref{eq:nlsd5} respectively into
Eq.~\eqref{eq:nlsd4} gives
\begin{equation}\label{eq:nlsd6}
\begin{split}
\int \delta\Psi_{\textrm{NL}}(\psi_{\mathbf{r'}}) \delta B(\psi_{\mathbf{r'}}) d\mathbf{r'} =&
\int \left| \int \nabla G_{\mathbf{r'},\mathbf{r''}} f_{\mathbf{r''}} 
\,d\mathbf{r''} \right|^{2}d\mathbf{r'}\\
&-\int\delta\psi_{\mathbf{r'}}f_{\mathbf{r'}}\,d\mathbf{r'}.
\end{split}
\end{equation}
Using Eq.~\eqref{eq:nlsd6} in Eq.~\eqref{eq:sd3} we obtain the following expression for the 
second variation of $\mathscr{K}_{\textrm{NL}}$ at its extremum:
\begin{equation}\label{eq:nlsd7}
\begin{split}
\delta^{2}\mathscr{K}_{\textrm{NL}}[\psi]\Big\vert_{\textrm{e}}=&
\,2\int \left| \int \nabla G_{\mathbf{r'},\mathbf{r''}} f_{\mathbf{r''}} 
d\mathbf{r''} \right|^{2}d\mathbf{r'}\\
&-2\int\delta\psi_{\mathbf{r'}}f_{\mathbf{r'}}d\mathbf{r'}
-\int \delta\psi_{\mathbf{r'}}\delta B(\psi_{\mathbf{r'}}) d\mathbf{r'}.
\end{split}
\end{equation}

We will now simplify the right hand side of Eq.~\eqref{eq:nlsd7} and in this regard expand 
the notation $f$ using Eq.~\eqref{eq:f} in the single integral above, to obtain
\begin{equation}
\begin{split}
\delta^{2}\mathscr{K}_{\textrm{NL}}[\psi]\Big\vert_{\textrm{e}}=&
\,2\int \left| \int \nabla G_{\mathbf{r'},\mathbf{r''}} f_{\mathbf{r''}} 
\,d\mathbf{r''} \right|^{2}d\mathbf{r'}\\
&- 2\int\delta\psi_{\mathbf{r'}}
\left(\delta B(\psi_{\mathbf{r'}}) -
\frac{1}{4\pi}\nabla^{2}\delta\psi_{\mathbf{r'}}\right) d\mathbf{r'}\\
&- \int \delta\psi_{\mathbf{r'}}\delta B(\psi_{\mathbf{r'}}) d\mathbf{r'},
\end{split}
\end{equation}
which further simplifies to
\begin{equation}\label{eq:nlsd8}
\begin{split}
\delta^{2}\mathscr{K}_{\textrm{NL}}[\psi]\Big\vert_{\textrm{e}}=&
\,2\int \left| \int \nabla G_{\mathbf{r'},\mathbf{r''}} f_{\mathbf{r''}} 
\,d\mathbf{r''} \right|^{2}d\mathbf{r'}\\
&- 3\int\delta\psi_{\mathbf{r'}}
\left(\delta B(\psi_{\mathbf{r'}}) -
\frac{2}{3}\frac{\nabla^{2}\delta\psi_{\mathbf{r'}}}{4\pi}\right) d\mathbf{r'}.
\end{split}
\end{equation}
Using Eq.~\eqref{eq:dB} in Eq.~\eqref{eq:nlsd8} and employing integration by parts leads to
\begin{equation}\label{eq:nlsd9}
\begin{split}
\delta^{2}\mathscr{K}_{\textrm{NL}}[\psi]\Big\vert_{\textrm{e}}=&
\,2\int \left| \int \nabla G_{\mathbf{r'},\mathbf{r''}} f_{\mathbf{r''}}
\,d\mathbf{r''} \right|^{2}d\mathbf{r'}\\
&+3\int\frac{\epsilon-2/3}{4\pi}
\left|\nabla\delta\psi_{\mathbf{r'}}\right|^{2}d\mathbf{r'}\\
&+ 3\int\beta\sum_{j=1}^{N}C_{j}q_{j}^{2}e^{-\beta q_{j}\psi_{\mathbf{r'}}}
\left(\delta\psi_{\mathbf{r'}}\right)^{2}d\mathbf{r'}.
\end{split}
\end{equation}
Inspecting each integral in the above equation
and noticing that concentrations are positive quantities and $\epsilon - 2/3 > 0$, 
it is clear that all the three integrals in Eq.~\eqref{eq:nlsd9} are non-negative. 
Moreover, since $\epsilon \ge 1$, the second integral in Eq.~\eqref{eq:nlsd9} is strictly positive, implying 
\begin{equation}
\delta^{2}\mathscr{K}_{\textrm{NL}}[\psi]\Big\vert_{\textrm{e}} > 0.
\end{equation}
The above inequality completes the proof that $\mathscr{K}_{\textrm{NL}}$ becomes a minimum at 
its extremum.


\begin{thebibliography}{26}%
\makeatletter
\providecommand \@ifxundefined [1]{%
 \@ifx{#1\undefined}
}%
\providecommand \@ifnum [1]{%
 \ifnum #1\expandafter \@firstoftwo
 \else \expandafter \@secondoftwo
 \fi
}%
\providecommand \@ifx [1]{%
 \ifx #1\expandafter \@firstoftwo
 \else \expandafter \@secondoftwo
 \fi
}%
\providecommand \natexlab [1]{#1}%
\providecommand \enquote  [1]{``#1''}%
\providecommand \bibnamefont  [1]{#1}%
\providecommand \bibfnamefont [1]{#1}%
\providecommand \citenamefont [1]{#1}%
\providecommand \href@noop [0]{\@secondoftwo}%
\providecommand \href [0]{\begingroup \@sanitize@url \@href}%
\providecommand \@href[1]{\@@startlink{#1}\@@href}%
\providecommand \@@href[1]{\endgroup#1\@@endlink}%
\providecommand \@sanitize@url [0]{\catcode `\\12\catcode `\$12\catcode
  `\&12\catcode `\#12\catcode `\^12\catcode `\_12\catcode `\%12\relax}%
\providecommand \@@startlink[1]{}%
\providecommand \@@endlink[0]{}%
\providecommand \url  [0]{\begingroup\@sanitize@url \@url }%
\providecommand \@url [1]{\endgroup\@href {#1}{\urlprefix }}%
\providecommand \urlprefix  [0]{URL }%
\providecommand \Eprint [0]{\href }%
\providecommand \doibase [0]{http://dx.doi.org/}%
\providecommand \selectlanguage [0]{\@gobble}%
\providecommand \bibinfo  [0]{\@secondoftwo}%
\providecommand \bibfield  [0]{\@secondoftwo}%
\providecommand \translation [1]{[#1]}%
\providecommand \BibitemOpen [0]{}%
\providecommand \bibitemStop [0]{}%
\providecommand \bibitemNoStop [0]{.\EOS\space}%
\providecommand \EOS [0]{\spacefactor3000\relax}%
\providecommand \BibitemShut  [1]{\csname bibitem#1\endcsname}%
\let\auto@bib@innerbib\@empty
\bibitem [{\citenamefont {Perutz}(1978)}]{perutz}%
  \BibitemOpen
  \bibfield  {author} {\bibinfo {author} {\bibfnamefont {M.}~\bibnamefont
  {Perutz}},\ }\href@noop {} {\bibfield  {journal} {\bibinfo  {journal}
  {Science}\ }\textbf {\bibinfo {volume} {201}},\ \bibinfo {pages} {1187}
  (\bibinfo {year} {1978})}\BibitemShut {NoStop}%
\bibitem [{\citenamefont {Honig}\ and\ \citenamefont {Nicholls}(1995)}]{honig}%
  \BibitemOpen
  \bibfield  {author} {\bibinfo {author} {\bibfnamefont {B.}~\bibnamefont
  {Honig}}\ and\ \bibinfo {author} {\bibfnamefont {A.}~\bibnamefont
  {Nicholls}},\ }\href@noop {} {\bibfield  {journal} {\bibinfo  {journal}
  {Science}\ }\textbf {\bibinfo {volume} {268}},\ \bibinfo {pages} {1144}
  (\bibinfo {year} {1995})}\BibitemShut {NoStop}%
\bibitem [{\citenamefont {Clapham}(2007)}]{clapham}%
  \BibitemOpen
  \bibfield  {author} {\bibinfo {author} {\bibfnamefont {D.~E.}\ \bibnamefont
  {Clapham}},\ }\href@noop {} {\bibfield  {journal} {\bibinfo  {journal}
  {Cell}\ }\textbf {\bibinfo {volume} {131}},\ \bibinfo {pages} {1047 }
  (\bibinfo {year} {2007})}\BibitemShut {NoStop}%
\bibitem [{\citenamefont {Levin}(2005)}]{levin1}%
  \BibitemOpen
  \bibfield  {author} {\bibinfo {author} {\bibfnamefont {Y.}~\bibnamefont
  {Levin}},\ }\href@noop {} {\bibfield  {journal} {\bibinfo  {journal} {Physica
  A: Statistical Mechanics and its Applications}\ }\textbf {\bibinfo {volume}
  {352}},\ \bibinfo {pages} {43 } (\bibinfo {year} {2005})}\BibitemShut
  {NoStop}%
\bibitem [{\citenamefont {Solis}\ \emph {et~al.}(2011)\citenamefont {Solis},
  \citenamefont {Vernizzi},\ and\ \citenamefont {Olvera de~la Cruz}}]{paco}%
  \BibitemOpen
  \bibfield  {author} {\bibinfo {author} {\bibfnamefont {F.~J.}\ \bibnamefont
  {Solis}}, \bibinfo {author} {\bibfnamefont {G.}~\bibnamefont {Vernizzi}}, \
  and\ \bibinfo {author} {\bibfnamefont {M.}~\bibnamefont {Olvera de~la
  Cruz}},\ }\href@noop {} {\bibfield  {journal} {\bibinfo  {journal} {Soft
  Matter}\ }\textbf {\bibinfo {volume} {7}},\ \bibinfo {pages} {1456} (\bibinfo
  {year} {2011})}\BibitemShut {NoStop}%
\bibitem [{\citenamefont {Vernizzi}\ and\ \citenamefont {Olvera de~la
  Cruz}(2007)}]{vernizzi}%
  \BibitemOpen
  \bibfield  {author} {\bibinfo {author} {\bibfnamefont {G.}~\bibnamefont
  {Vernizzi}}\ and\ \bibinfo {author} {\bibfnamefont {M.}~\bibnamefont {Olvera
  de~la Cruz}},\ }\href@noop {} {\bibfield  {journal} {\bibinfo  {journal}
  {Proceedings of the National Academy of Sciences}\ }\textbf {\bibinfo
  {volume} {104}},\ \bibinfo {pages} {18382} (\bibinfo {year}
  {2007})}\BibitemShut {NoStop}%
\bibitem [{\citenamefont {L\"owen}\ \emph {et~al.}(1992)\citenamefont
  {L\"owen}, \citenamefont {Madden},\ and\ \citenamefont {Hansen}}]{lowen}%
  \BibitemOpen
  \bibfield  {author} {\bibinfo {author} {\bibfnamefont {H.}~\bibnamefont
  {L\"owen}}, \bibinfo {author} {\bibfnamefont {P.~A.}\ \bibnamefont {Madden}},
  \ and\ \bibinfo {author} {\bibfnamefont {J.-P.}\ \bibnamefont {Hansen}},\
  }\href@noop {} {\bibfield  {journal} {\bibinfo  {journal} {Phys. Rev. Lett.}\
  }\textbf {\bibinfo {volume} {68}},\ \bibinfo {pages} {1081} (\bibinfo {year}
  {1992})}\BibitemShut {NoStop}%
\bibitem [{\citenamefont {Andelman}(1995)}]{andelman}%
  \BibitemOpen
  \bibfield  {author} {\bibinfo {author} {\bibfnamefont {D.}~\bibnamefont
  {Andelman}},\ }\href@noop {} {\bibfield  {journal} {\bibinfo  {journal}
  {Handbook of biological physics}\ }\textbf {\bibinfo {volume} {1}},\ \bibinfo
  {pages} {603} (\bibinfo {year} {1995})}\BibitemShut {NoStop}%
\bibitem [{\citenamefont {Sharp}\ and\ \citenamefont {Honig}(1990)}]{honig1}%
  \BibitemOpen
  \bibfield  {author} {\bibinfo {author} {\bibfnamefont {K.~A.}\ \bibnamefont
  {Sharp}}\ and\ \bibinfo {author} {\bibfnamefont {B.}~\bibnamefont {Honig}},\
  }\href@noop {} {\bibfield  {journal} {\bibinfo  {journal} {The Journal of
  Physical Chemistry}\ }\textbf {\bibinfo {volume} {94}},\ \bibinfo {pages}
  {7684} (\bibinfo {year} {1990})}\BibitemShut {NoStop}%
\bibitem [{\citenamefont {Reiner}\ and\ \citenamefont {Radke}(1990)}]{radke}%
  \BibitemOpen
  \bibfield  {author} {\bibinfo {author} {\bibfnamefont {E.~S.}\ \bibnamefont
  {Reiner}}\ and\ \bibinfo {author} {\bibfnamefont {C.~J.}\ \bibnamefont
  {Radke}},\ }\href@noop {} {\bibfield  {journal} {\bibinfo  {journal} {J.
  Chem. Soc.{,} Faraday Trans.}\ }\textbf {\bibinfo {volume} {86}},\ \bibinfo
  {pages} {3901} (\bibinfo {year} {1990})}\BibitemShut {NoStop}%
\bibitem [{\citenamefont {Fogolari}\ and\ \citenamefont
  {Briggs}(1997)}]{fogolari}%
  \BibitemOpen
  \bibfield  {author} {\bibinfo {author} {\bibfnamefont {F.}~\bibnamefont
  {Fogolari}}\ and\ \bibinfo {author} {\bibfnamefont {J.~M.}\ \bibnamefont
  {Briggs}},\ }\href@noop {} {\bibfield  {journal} {\bibinfo  {journal}
  {Chemical Physics Letters}\ }\textbf {\bibinfo {volume} {281}},\ \bibinfo
  {pages} {135 } (\bibinfo {year} {1997})}\BibitemShut {NoStop}%
\bibitem [{\citenamefont {Che}\ \emph {et~al.}(2008)\citenamefont {Che},
  \citenamefont {Dzubiella}, \citenamefont {Li},\ and\ \citenamefont
  {McCammon}}]{che}%
  \BibitemOpen
  \bibfield  {author} {\bibinfo {author} {\bibfnamefont {J.}~\bibnamefont
  {Che}}, \bibinfo {author} {\bibfnamefont {J.}~\bibnamefont {Dzubiella}},
  \bibinfo {author} {\bibfnamefont {B.}~\bibnamefont {Li}}, \ and\ \bibinfo
  {author} {\bibfnamefont {J.~A.}\ \bibnamefont {McCammon}},\ }\href@noop {}
  {\bibfield  {journal} {\bibinfo  {journal} {The Journal of Physical Chemistry
  B}\ }\textbf {\bibinfo {volume} {112}},\ \bibinfo {pages} {3058} (\bibinfo
  {year} {2008})},\ \bibinfo {note} {pMID: 18275182}\BibitemShut {NoStop}%
\bibitem [{\citenamefont {Maggs}(2012)}]{maggs1}%
  \BibitemOpen
  \bibfield  {author} {\bibinfo {author} {\bibfnamefont {A.~C.}\ \bibnamefont
  {Maggs}},\ }\href@noop {} {\bibfield  {journal} {\bibinfo  {journal} {EPL
  (Europhysics Letters)}\ }\textbf {\bibinfo {volume} {98}},\ \bibinfo {pages}
  {16012} (\bibinfo {year} {2012})}\BibitemShut {NoStop}%
\bibitem [{\citenamefont {Kung}\ \emph {et~al.}(2009)\citenamefont {Kung},
  \citenamefont {Solis},\ and\ \citenamefont {Olvera de~la Cruz}}]{kung}%
  \BibitemOpen
  \bibfield  {author} {\bibinfo {author} {\bibfnamefont {W.}~\bibnamefont
  {Kung}}, \bibinfo {author} {\bibfnamefont {F.~J.}\ \bibnamefont {Solis}}, \
  and\ \bibinfo {author} {\bibfnamefont {M.}~\bibnamefont {Olvera de~la
  Cruz}},\ }\href@noop {} {\bibfield  {journal} {\bibinfo  {journal} {The
  Journal of Chemical Physics}\ }\textbf {\bibinfo {volume} {130}},\ \bibinfo
  {eid} {044502} (\bibinfo {year} {2009})}\BibitemShut {NoStop}%
\bibitem [{\citenamefont {Jha}\ \emph {et~al.}(2009)\citenamefont {Jha},
  \citenamefont {Solis}, \citenamefont {de~Pablo},\ and\ \citenamefont {Olvera
  de~la Cruz}}]{prateek}%
  \BibitemOpen
  \bibfield  {author} {\bibinfo {author} {\bibfnamefont {P.~K.}\ \bibnamefont
  {Jha}}, \bibinfo {author} {\bibfnamefont {F.~J.}\ \bibnamefont {Solis}},
  \bibinfo {author} {\bibfnamefont {J.~J.}\ \bibnamefont {de~Pablo}}, \ and\
  \bibinfo {author} {\bibfnamefont {M.}~\bibnamefont {Olvera de~la Cruz}},\
  }\href@noop {} {\bibfield  {journal} {\bibinfo  {journal} {Macromolecules}\
  }\textbf {\bibinfo {volume} {42}},\ \bibinfo {pages} {6284} (\bibinfo {year}
  {2009})}\BibitemShut {NoStop}%
\bibitem [{\citenamefont {van Roij}\ and\ \citenamefont
  {Hansen}(1997)}]{van_roij}%
  \BibitemOpen
  \bibfield  {author} {\bibinfo {author} {\bibfnamefont {R.}~\bibnamefont {van
  Roij}}\ and\ \bibinfo {author} {\bibfnamefont {J.-P.}\ \bibnamefont
  {Hansen}},\ }\href@noop {} {\bibfield  {journal} {\bibinfo  {journal} {Phys.
  Rev. Lett.}\ }\textbf {\bibinfo {volume} {79}},\ \bibinfo {pages} {3082}
  (\bibinfo {year} {1997})}\BibitemShut {NoStop}%
\bibitem [{\citenamefont {Knott}\ and\ \citenamefont {Ford}(2001)}]{knott}%
  \BibitemOpen
  \bibfield  {author} {\bibinfo {author} {\bibfnamefont {M.}~\bibnamefont
  {Knott}}\ and\ \bibinfo {author} {\bibfnamefont {I.~J.}\ \bibnamefont
  {Ford}},\ }\href@noop {} {\bibfield  {journal} {\bibinfo  {journal} {Phys.
  Rev. E}\ }\textbf {\bibinfo {volume} {63}},\ \bibinfo {pages} {031403}
  (\bibinfo {year} {2001})}\BibitemShut {NoStop}%
\bibitem [{\citenamefont {Rottler}\ and\ \citenamefont
  {Maggs}(2004)}]{rottler-maggs}%
  \BibitemOpen
  \bibfield  {author} {\bibinfo {author} {\bibfnamefont {J.}~\bibnamefont
  {Rottler}}\ and\ \bibinfo {author} {\bibfnamefont {A.~C.}\ \bibnamefont
  {Maggs}},\ }\href@noop {} {\bibfield  {journal} {\bibinfo  {journal} {Phys.
  Rev. Lett.}\ }\textbf {\bibinfo {volume} {93}},\ \bibinfo {pages} {170201}
  (\bibinfo {year} {2004})}\BibitemShut {NoStop}%
\bibitem [{\citenamefont {Car}\ and\ \citenamefont
  {Parrinello}(1985)}]{car-parrinello}%
  \BibitemOpen
  \bibfield  {author} {\bibinfo {author} {\bibfnamefont {R.}~\bibnamefont
  {Car}}\ and\ \bibinfo {author} {\bibfnamefont {M.}~\bibnamefont
  {Parrinello}},\ }\href@noop {} {\bibfield  {journal} {\bibinfo  {journal}
  {Phys. Rev. Lett.}\ }\textbf {\bibinfo {volume} {55}},\ \bibinfo {pages}
  {2471} (\bibinfo {year} {1985})}\BibitemShut {NoStop}%
\bibitem [{\citenamefont {Jadhao}\ \emph {et~al.}(2012)\citenamefont {Jadhao},
  \citenamefont {Solis},\ and\ \citenamefont {Olvera de~la Cruz}}]{jso1}%
  \BibitemOpen
  \bibfield  {author} {\bibinfo {author} {\bibfnamefont {V.}~\bibnamefont
  {Jadhao}}, \bibinfo {author} {\bibfnamefont {F.~J.}\ \bibnamefont {Solis}}, \
  and\ \bibinfo {author} {\bibfnamefont {M.}~\bibnamefont {Olvera de~la
  Cruz}},\ }\href@noop {} {\bibfield  {journal} {\bibinfo  {journal} {Phys.
  Rev. Lett.}\ }\textbf {\bibinfo {volume} {109}},\ \bibinfo {pages} {223905}
  (\bibinfo {year} {2012})}\BibitemShut {NoStop}%
\bibitem [{\citenamefont {Jadhao}\ \emph {et~al.}(2013)\citenamefont {Jadhao},
  \citenamefont {Solis},\ and\ \citenamefont {Olvera de~la Cruz}}]{jso2}%
  \BibitemOpen
  \bibfield  {author} {\bibinfo {author} {\bibfnamefont {V.}~\bibnamefont
  {Jadhao}}, \bibinfo {author} {\bibfnamefont {F.~J.}\ \bibnamefont {Solis}}, \
  and\ \bibinfo {author} {\bibfnamefont {M.}~\bibnamefont {Olvera de~la
  Cruz}},\ }\href@noop {} {\bibfield  {journal} {\bibinfo  {journal} {The
  Journal of Chemical Physics}\ }\textbf {\bibinfo {volume} {138}},\ \bibinfo
  {eid} {054119} (\bibinfo {year} {2013})}\BibitemShut {NoStop}%
\bibitem [{\citenamefont {Remler}\ and\ \citenamefont {Madden}(1990)}]{madden}%
  \BibitemOpen
  \bibfield  {author} {\bibinfo {author} {\bibfnamefont {D.}~\bibnamefont
  {Remler}}\ and\ \bibinfo {author} {\bibfnamefont {P.}~\bibnamefont
  {Madden}},\ }\href@noop {} {\bibfield  {journal} {\bibinfo  {journal}
  {Molecular Physics}\ }\textbf {\bibinfo {volume} {70}},\ \bibinfo {pages}
  {921} (\bibinfo {year} {1990})}\BibitemShut {NoStop}%
\bibitem [{\citenamefont {Baptista}\ \emph {et~al.}(2009)\citenamefont
  {Baptista}, \citenamefont {Schmitz},\ and\ \citenamefont
  {D\"unweg}}]{baptista}%
  \BibitemOpen
  \bibfield  {author} {\bibinfo {author} {\bibfnamefont {M.}~\bibnamefont
  {Baptista}}, \bibinfo {author} {\bibfnamefont {R.}~\bibnamefont {Schmitz}}, \
  and\ \bibinfo {author} {\bibfnamefont {B.}~\bibnamefont {D\"unweg}},\
  }\href@noop {} {\bibfield  {journal} {\bibinfo  {journal} {Phys. Rev. E}\
  }\textbf {\bibinfo {volume} {80}},\ \bibinfo {pages} {016705} (\bibinfo
  {year} {2009})}\BibitemShut {NoStop}%
\bibitem [{\citenamefont {Maggs}(2004)}]{maggs_cite}%
  \BibitemOpen
  \bibfield  {author} {\bibinfo {author} {\bibfnamefont {A.~C.}\ \bibnamefont
  {Maggs}},\ }\href@noop {} {\bibfield  {journal} {\bibinfo  {journal} {The
  Journal of Chemical Physics}\ }\textbf {\bibinfo {volume} {120}},\ \bibinfo
  {pages} {3108} (\bibinfo {year} {2004})}\BibitemShut {NoStop}%
\bibitem [{\citenamefont {Jackson}(1999)}]{jackson}%
  \BibitemOpen
  \bibfield  {author} {\bibinfo {author} {\bibfnamefont {J.~D.}\ \bibnamefont
  {Jackson}},\ }\href@noop {} {\emph {\bibinfo {title} {Classical
  Electrodynamics}}},\ \bibinfo {edition} {3rd}\ ed.\ (\bibinfo  {publisher}
  {Wiley, New York},\ \bibinfo {year} {1999})\BibitemShut {NoStop}%
\bibitem [{\citenamefont {Schwinger}\ \emph {et~al.}(1998)\citenamefont
  {Schwinger}, \citenamefont {Deraad}, \citenamefont {Milton}, \citenamefont
  {Tsai},\ and\ \citenamefont {Norton}}]{schwinger}%
  \BibitemOpen
  \bibfield  {author} {\bibinfo {author} {\bibfnamefont {J.}~\bibnamefont
  {Schwinger}}, \bibinfo {author} {\bibfnamefont {L.}~\bibnamefont {Deraad}},
  \bibinfo {author} {\bibfnamefont {K.}~\bibnamefont {Milton}}, \bibinfo
  {author} {\bibfnamefont {W.}~\bibnamefont {Tsai}}, \ and\ \bibinfo {author}
  {\bibfnamefont {J.}~\bibnamefont {Norton}},\ }\href@noop {} {\emph {\bibinfo
  {title} {Classical Electrodynamics}}},\ Advanced book program\ (\bibinfo
  {publisher} {Westview Press},\ \bibinfo {year} {1998})\BibitemShut {NoStop}%
\end{thebibliography}
\end{document}